\documentstyle[12pt]{article}
\def \o{\omega}
\def \t{\tilde}
\def \i{\int}
\def \p{\partial}
\def \m{m_{pl}^2}
\def \s{\sigma}
\def \n{\nonumber}
\def \d{\delta}
\def \r{\sqrt{12\pi}}
\def \e{\epsilon}
\def \G{\Gamma}
\def \L{\Lambda}
\def \b{\Biggl({b \over b_0}\Biggr)}
\def \be{\begin{eqnarray}}
\def \en{\end{eqnarray}}
\def \th{\theta}
\begin{document}
\baselineskip=20pt
\centerline{\Large{\bf CONSTRAINTS ON HIGHER DIMENSIONAL MODELS}}
\vskip 0.2in
\centerline{\Large{\bf FOR VIABLE EXTENDED INFLATION}}

\vskip 0.5in

\centerline{A. S. Majumdar\footnote{e-mail:archan@bose.ernet.in}}
\vskip 0.2in

\centerline{S.N.Bose National Centre for Basic Sciences}
\vskip 0.1in
\centerline{Block JD, Sector III, Salt Lake, Calcutta 700091, India}

\vskip 0.5in

{\bf Abstract}

We consider two kinds of higher dimensional models
which upon dimensional reduction lead to
Jordan-Brans-Dicke type effective actions in four
dimensions with the scale factor of the extra
dimensions playing the role of the JBD field.
These models are characterized by the potential
for the JBD field which arises from the process of
dimensional reduction, and by the coupling of the
inflaton sector with the JBD field in the Jordan
frame.
Taking into account the fact that these models
allow the possibility of enough inflation and
dynamical compactification of the extra
dimensions, we examine in the context of these
models the other conditions which
need to be satisfied for a viable scenario of
extended inflation. We find that the requirements of conforming
to general relativity at the present epoch, and
producing suitable bubble spectrums during
inflation lead to  constraints on the
allowed values taken by the parameters of these
models. A model with a ten dimensional JBD field is able to satisfy
the condition for appropriate density
perturbations viewed in the conformal Einstein
frame, with a stringent restriction on the initial
value taken by the scale factor of the extra
dimensions.

\vskip 0.2in
PACS number(s): 98.80.Cq, 11.10.Kk

\pagebreak

\section{Introduction}

Gravitational field theories in which the spatial dimensions exceed
the usual three, have been widely studied in the last several
years~[1]. A primary reason for this interest is the fact that
several of these higher dimensional models are obtained in the low
energy (point particle) limit of string theories~[2]. The former
are used to investigate the cosmological consequences of the latter
models, and several aspects of dilaton cosmology have been revealed
recently~[3]. A key common feature of higher dimensional models is
that upon dimensional reduction they lead to scalar-tensor
theories~[4] of gravity in the four spacetime dimensions, of which
the Jordan-Brans-Dicke~[5] (JBD) model is one particular example.

It is well known that scalar-tensor theories in four dimensions act
as candidates for extended inflation (EI)~[6]. The scenario of EI
restores the spirit of `old' inflation~[7] in the sense that
inflation is driven by the vacuum energy of a scalar field
(inflaton) trapped in a metastable state, which subsequently
tunnels out of the potential barrier through a predominantly first
order phase transition by the formation of bubbles of true
vacuum~[8]. Although there exist several models which can implement
EI (to the extent of alleviating the typical problems of other
versions of inflation like `old' inflation~[7], `new' and `chaotic'
inflation~[9]), this scheme encounters some of its own
characteristic problems such as the `$\omega$-problem', which have created
obstacles towards building a realistic model.

The `$\omega$-problem' of EI is described in a simple fashion in the
context of the JBD theory where a scalar filed (JBD field) with
coupling parameter `$\omega$' takes the place of the gravitational
constant. The maximum departure from general relativity allowed by
present observations forces the constraint $\omega > 500$ at the
present epoch~[10]. On
the other hand, it is required for the phase transition of the
inflaton field to proceed in such a manner that the nucleation rate
of earliest bubbles which have the potential to cause large
anisotropies in the microwave background, should be suppressed. The
desiradatum for a suitable bubble distribution restrains $\omega$
to be $\omega < 25$ during inflation~[11]. These two bounds on
$\omega$ are incompatible for the case of the simplest JBD model
where $\omega$ is a constant. Although it is possible to construct
more sophisticated models with variable $\o$, or with potentials
for the JBD field~[12,13], the difficulty of implementing appropriate
density perturbations remains in such models. It is widely believed
that quantum fluctuations of the scalar fields during inflation leading
to density perturbations should be able to provide the seeds of
large scale structure formation in the universe. The analysis of
the COBE results~[14] on CMBR anisotropy and various other large
scale structure observations have led to the imposition of
stringent constraints on the potential used for inflation~[15].

Recently, Green and Liddle~[16] have summarized all the conditions
required for the construction of successful EI models. They have
shown the incompatibility of a large variety of existing EI models
(except for some specially contrived ones)
in meeting these requirements. Nevertheless, as we shall argue
below, there exist a few models of EI obtained from
compactification of higher dimensional theories, which lie outside
the general category of models considered in [16]. For example, it
has been observed for a model with a nonminimally coupled inflaton
field in higher dimensions~[17], that a four dimensional EI
scenario emerges where the effective JBD field  after the
completion of inflation is anchored in a potential which follows
naturally from the higher dimensional action. The rate of bubble
nucleation is time dependent, a feature which is generic to such
higher dimensional models~[17,18]. Another type of Kaluza-Klein
models can lead to four dimensional lagrangians with variable $\o$.
Such an example was provided in [19] where together with enough
inflation, stable compactification of the extra spatial dimensions
was achieved by considering the dynamics in the conformal Einstein
frame. The next step is to study to what extent these models
[17-19] are capable of accommodating the recently formulated
constraints from density perturbations~[15].

An unsatisfactory feature of such higher dimensional models is the
presence of several `free' parameters, e.g., scale of curvature of
the extra dimensions, strength of the cosmological term, etc. It is
expected that the various considerations for implementing a
successful EI scenario would impose restrictions on the choice of
values for these parameters. With this aim in mind, in this paper
we study the dynamics of EI in two types of dimensionally reduced
Kaluza-Klein models. In Section II we consider a model with
nonminimal coupling for the inflaton field in the conformally
transformed Einstein frame. Solutions for this model was earlier
worked out in the Jordan frame~[17]. It is true that a conformal
transformation of the spacetime metric could in principle lead to
certain differences in the dynamics~[20]. However, for the purpose
of the present analysis, such differences can be ignored~[16] by
requiring that the models considered are rendered sufficiently
close to the limit of general relativity~[10] after the epoch of
inflation. We analyze all the required conditions for succesful EI
in context of the specific model.
We find that although the model of Section II obeys most of the
other criteria for a suitable range of parameters, it is incapable
of meeting the requirements for successful density perturbations.
In Section III, we study a higher dimensional JBD model~[19] in the
Einstein frame. Our analysis shows that this model can accommodate
all the conditions by imposing tight constraints on the parameters.
We present a brief summary of our results in Section IV.

\section{Model with nonminimally coupled inflaton field in
$4+D$ dimensions}

Before introducing our model, it needs to be stated that the model
with the standard coupling for the inflaton in ten ($D=6$)
dimensions, which was analyzed in [18], fails to simultaneously
provide enough inflation for the four dimensional scale factor and
achieve compactification for the extra dimensions. The utility of
nonminimal coupling for scalar fields in higher dimensions has been
noted in [21] where it was observed that the choice of a certain
range of values for the coupling parameter could prohibit the
isotropic expansion of all the $4+D$ dimensions. Our model in $4+D$
dimensions is given by the action
\begin{eqnarray}
{\t S} = \int d^{10}z (-{\t g})^{1/2}\bigl[ -{{\t R}\over 16\pi{\t
G}} + {1\over 2}{\t g}^{MN} \p_M {\t \chi} \p_N {\t \chi} - \xi {\t
R} ({\t \chi}^2 - {\t \chi}_0^2) - {\t U}({\t \chi})\bigr]
\end{eqnarray}
Tildes are used throughout to describe $4+D$ dimensional
quantities. It is assumed that the inflaton field ${\t \chi}$ is
anchored in a metastable state, from where it tunnels out to the
true vacuum through the nucleation of bubbles. The line element in
$4+D$ dimensions is chosen of the form
\begin{eqnarray}
d{\t s}^2 = dt^2 - a^2(t)d{\Omega}^2_3 - b^2(t)d{\Omega}^2_D
\end{eqnarray}
where $d{\Omega}^2_3$ is the line element of a maximally symmetric
$3$-space with scale factor $a(t)$ and $d{\Omega}^2_D$ corresponds
to a $D$-sphere with scale factor $b(t)$.

We follow the usual prescription of dimensional reduction~[1,17,18]
with the following definitions
\begin{eqnarray}
{\Omega}_D = {2\pi^{(D+1)/2} \over \Gamma (D+1)/2} \nonumber \\
{{\Omega}_D b_0^D \over {\t G}} = m_{pl}^2 \nonumber \\
\alpha = {D(D-1) \over b_0^2} \Biggl( {\m \over 16\pi}\Biggr)^{2/D}
\nonumber \\
\s = ({\Omega}_D b_0^D)^{1/2} \chi \nonumber \\
V(\s) = ({\Omega}_D b_0^D) U(\chi) \nonumber \\
V_0 = {8\pi \over \m} V(\s = 0) \nonumber \\
\d = 1 - {16\pi \over \m} \xi \s_0^2 \n \\
\Phi = {(b/b_0)^D \m \over 16\pi}
\end{eqnarray}
where $b_0$ is a parameter with dimensions of length. $\s$ is the
inflaton field in four dimensions which undergoes phase transition
from its false vacuum value ($\s = 0$) to the true vacuum defined
at ($\s = \s_0$). We consider the ten dimensional ($D=6$) case of
(1) which upon dimensional reduction to four dimensions yields an
effective action of the JBD type given by
\begin{eqnarray}
S = \i d^4x (-g)^{1/2} \bigl[ -\d \Phi R - {5\over 6} \d g^{\mu\nu}
{\p_{\mu}\Phi \p_{\nu}\Phi \over \Phi} + \d \alpha \Phi^{2/3}
-2\Phi V_0 \Bigr]
\end{eqnarray}

At this stage a comparison with the action in the Jordan frame
representing the general class of models considered in [16] is in
order. The main differences are (i) the kinetic term of the JBD
field $\Phi$ (representing the scale factor of the six dimensional
internal manifold) comes with an opposite sign, (ii) a potential
for $\Phi$ occurs naturally and cannot be set to zero unlike as
done for the subsequent analysis in the Einstein frame in [16],
(iii) the inflaton sector gets coupled to the JBD field $\Phi$ in
the Jordan frame itself, and (iv) the last two terms in (4)
represent the total effective potential of the inflaton field in
the Jordan frame. It is clear that the analysis of [16] does not
exhaust all the possibilities provided by models of the type (4).

Equations of motion following from (4) were solved numerically in
[17] where it was seen that inflation accrues for the scale
factor $a(t)$, whereas the field $\Phi$ ($\sim$ internal scale
factor $b(t)$) rolls down the hill of its potential $V(\Phi)$. An
extra Maxwell type field was introduced to maintain stability of
compactification at a nonvanishing value of $b(t)$, since it
induces a local minimum of $V(\Phi)$ at the corresponding value.
The above analysis was carried out in the Jordan frame.
Nevertheless, a conformal transformation to the Einstein frame not
only simplifies the gravitational sector of the theory, but is also
essential for studying some key details of EI~[16,22]. Henceforth
we shall concentrate on these aspects in the conformal Einstein
frame.

We define the Einstein frame metric ${\bar g}_{\mu\nu}$ as
\begin{eqnarray}
{\bar g}_{\mu\nu} = {\Omega}^{-2} g_{\mu\nu} \n \\
{\Omega}^{-2} = {16\pi \d \Phi \over \m}
\end{eqnarray}
and a new scalar field $y$ as
\begin{eqnarray}
y = {m_{pl} \over \r} ln\bigl(\Phi/\Phi_0\bigr)
\end{eqnarray}
where $\Phi_0$ is a parameter having dimensions of $(mass)^2$.
Using (5) and (6) in (1), the action in the Einstein frame
takes the form
\begin{eqnarray}
{\bar S} = \i d^4x (-{\bar g})^{1/2} \Biggl[- {\m \over 16\pi}
{\bar R} + {1\over 2} {\bar g}^{\mu\nu} \p_{\mu} y \p_{\nu} y -
V(y)\Biggr]
\end{eqnarray}
with $V(y)$ given by
\begin{eqnarray}
V(y) = \Biggl({\m \over 16\pi}\Biggr)^2 \Biggl[2{V_0 \over \d^2 \Phi_0}
exp\Biggl(- {\r y \over m_{pl}}\Biggr) - {\alpha
\over \d \Phi_0^{4/3}} exp\Biggl(-{4\over 3} {\r y
\over m_{pl}}\Biggr)\Biggr]
\end{eqnarray}

Having defined the model in the Einstein frame we shall now analyze
the various criteria which need to be satisfied for a viable
scenario of extended inflation. These conditions can be stated as
(a) recovering general relativity after the end of inflation within
the freedom allowed by present experiments, (b) reproducing the
correct strength of the gravitational coupling, (c) obtaining a
bubble spectrum which agrees with CMBR isotropy, and (d) generating
density perturbations that are compatible with large scale
structure observations. Let us now consider the specific details of
these separate desiradata in the context of the present model.

To begin with, note that the potential for this model V(y) (8)
has no stable minimum, thus allowing the field $y$ to role down
unhindered. (See Fig.1 where the dimensionless function $U(y) = (16\pi
b_0^2/\m)V(y)$ is plotted versus $y/m_{pl}$ for certain typical values of
the parameters.) As stated earlier, it is known~[17,18] that this
defect can be rectified by the addition of a Maxwell-type term in
the higher dimensional action (1). Such a term plays the role of
introducing a local minimum in $V(y)$ at a small  value
of $y$, but has negligible effect on the dynamical evolution for
large values of $y$. A similar role can be performed by the
introduction of a Casimir term in higher dimensions, as we shall do
in Section III. However, to keep the analysis as simple as possible
in the present case, one can assume that $y$ can be anchored at a
local minimum of the potential, without explicitely writing the
required extra term in the Lagrangian. With this assumption,
general relativity is exactly recovered once $y$ stops rolling
down. From (4) it is clear that reproducing the present value of
the gravitational coupling enforces the condition
\begin{eqnarray}
y_{now} = {m_{pl} \over \r} ln\Biggl({\m \over
16\pi\d\Phi_0}\Biggr)
\end{eqnarray}
where $y_{now}$ denotes the present value of $y$. If $y$ is
anchored at the local minimum of $V(y)$ at the end of inflation,
then $y_{now} = y_{end}$ ($y_{end}$ denotes the value of $y$ at the
end of inflation). Since $y_{end}$ is approximately zero, the value
of $\Phi_0$ is constrained to be
\begin{eqnarray}
\Phi_0 \simeq  {\m \over 16\pi\d}
\end{eqnarray}

In order to examine the constraints from bubble spectrum and
density perturbations, it is customary to define, in the slow role
approximation, the parameters $\e(y)$ and $\eta(y)$ [8,15]
associated with the potential $V(y)$ as
\begin{eqnarray}
\e(y) = {\m \over 16\pi} \Biggl({V' \over V}\Biggr)^2 \n \\
\eta(y) = {\m \over 8\pi} {V'' \over V}
\end{eqnarray}
with the primes indicating derivatives with respect to $y$. The
number of e-foldings of inflation between two values of $y$ is
approximately given by
\begin{eqnarray}
N(y_1, y_2) \simeq -{8\pi \over \m} \i_{y_1}^{y_2} {V \over V'}dy
\end{eqnarray}
For the present model obtaining the expressions of $V$ and $V'$
from (8), it turns out that the right hand side of (12) is of a
rather simple form. The expression for $N$ is given by
\begin{eqnarray}
N(y_1, y_2) \simeq {\r \over 2m_{pl}} (y_2 - y_1)
\end{eqnarray}

The progress of phase transition of the inflaton field in an
expanding background is determined by the quantity
\begin{eqnarray}
E(t) = {\Gamma(t) \over H^4(t)}
\end{eqnarray}
which measures the percentage of false vacuum occupied by bubbles
of true vacuum ($E=1$ signals the end of phase transition). $H$ is
the Hubble parameter and $\Gamma$ is the nucleation rate of true
vacuum bubbles per unit time per unit volume. It needs to be
emphasized that unlike some of the less complicated EI
scenarios~[6,11,16], $\G(t)$ is not a constant here, but varies
according to the time evolution of $y$. This feature is generic to
dimensionally reduced Kaluza-Klein models~[17,18,19] since the JBD
field in four dimensions gets coupled to the inflaton sector in the
Jordan frame itself. Although the calculation of nucleation rate in
the presence of time dependent fields is a complicated
problem~[23], in the limit of weak gravity, it is possible to write
down a closed form expression for $\G(y)$ [17,18,24] which for the
present model reduces to
\begin{eqnarray}
\G(y) = A_0 exp\Biggl[-B_0 exp\Biggl({\r y \over m_{pl}}\Biggr) +
{2\r y \over m_{pl}}\Biggr]
\end{eqnarray}
where $A_0 \sim \s_0^4$, and $B_0$ is the flat space bounce action.

The time evolution of $\G(y)$ was computed in [18] using solutions
for $y$ in the Jordan frame. It was found to be extremely favorable
for the desired bubble distribution as the early (large $y$)
formation of bubbles is exponentially suppressed. This point can be
understood better by considering the value of various quantities at
an era when bubble formation has just begun to take place. At this
instant (about 55 e-foldings from the end of inflation), we call
the value of $y$ as $y_{55}$. In order that the unthermalyzed
bubbles from this stage do not inflate up to lead to unobserved
anisotropies in the CMBR, an upper bound is placed on the value of
$E$ at this epoch ($E_{55}$) [16,25]
\begin{eqnarray}
E_{55} = {\G_{55} \over H_{55}} < 10^{-5}
\end{eqnarray}

To calculate the value of $\G_{55}$ (15) for our model one needs
to first know the possible values taken by $y_{55}$. This can be
easily fixed from (13) by substituting $y_2 = y_{55}$ and $y_1 =
y_{end} \simeq 0$, to obtain
\begin{eqnarray}
y_{55} \simeq {110 m_{pl} \over \r}
\end{eqnarray}
Now substituting (17) in (15) and using the fact that the
Jordan frame Hubble parameter is approximately proportional to the
fourth root of the Einstein frame potential~[16], the bound (16)
translates into the condition (after the normalyzation $E_{end} =1$)
\begin{eqnarray}
E_{55} = {\G_{55} \over \G_{end}} \Biggl({H_{end} \over
H_{55}}\Biggr)^4 \simeq exp\biggl[- B_0 exp(110)\biggr]{V_{end}
\over V_{55}} < 10^{-5}
\end{eqnarray}
By substituting the values of the various quantities from (8),
(10) and (17), it can be seen that (18) is easily satisfied.
The necessity of a suitable bubble distribution does not impose any
additional constraints on the parameters of this model.

The final check of testing the viability of this model comes from
the requirement of generating density perturbations conforming to
observations. This part of the analysis depends upon the specific
cosmological model used (i.e., contributions to the energy density
from different forms of dark matter, cosmological constant, etc.)
[8]. Without going into these details, we adopt the criteria
developed by Liddle et al.[15,16] which places constraints on the
parameters $\e$ and $\eta$ (11) at the epoch of $55$ e-foldings
from the end of inflation. It is required that
\begin{eqnarray}
4\e_{55} - \eta_{55} < 0.2 \n \\
\eta_{55} - \e_{55} < 0.1
\end{eqnarray}

To see whether the above two conditions are satisfied, we first
substitute the values of $y_{55}$ (17) and $\Phi_0$ (10) into
$V(y)$ and its derivatives obtained from (8). Then, upon using
(11) we find that both $\e_{55}$ and $\eta_{55}$ can be made to
possess values of $O(10^{-1})$ by demanding that
\begin{eqnarray}
{b_0^2 V_0 \over \d} \simeq 10^{-15}
\end{eqnarray}
This constraint on the parameters $b_0$ (scale of the internal
manifold), $V_0$ (3), and $\d$ (nonminimal coupling parameter) is
in fact a requirement for enough inflation ($\e, \eta << 1$). If
the phase transition for the inflaton field is assumed to take
place around the GUT scale, a particular choice which satisfies
(21) is $b_0 \sim O(10(m_{pl})^{-1}), \d \sim O(1)$. However, a
closer scrutiny of the relations (19) by substituting the
appropriate numbers, shows that both of them can never be satisfied
together. The reason for this model failing to generate appropriate
density perturbations is because of the form of the potential
(8). Any tuning of the parameters is unable to salvage the
scenario.

\section{Higher dimensional JBD model}

The action in $4+D$ dimensions is
\begin{eqnarray}
{\t S} = {1\over 16\pi} \i d^{4+D}z (-{\t g})^{1/2} \Biggl[-{\t
\Phi}{\t R} + {\t \o} {\t g}^{MN} {\p_M{\t \Phi} \p_N {\t \Phi}
\over {\t \Phi}} - {\t \L} + {\cal L}({\t \chi})\Biggr]
\end{eqnarray}
where ${\t \Phi}$ and ${\t \L}$ are the $4+D$ dimensional JBD field
and cosmological constant respectively. ${\cal L}({\t \chi})$
is the Lagrangian for the inflaton field which is caught in the
metastable state of its potential. The line element is again
assumed to be of the form (2). Employing the usual procedure of
dimensional reduction one obtains the four dimensional action in
the Jordan frame
\begin{eqnarray}
S = \i d^4x (-g)^{1/2} \Biggl[-\Phi \Biggl({b \over b_0}\Biggr)^D R
- \Phi \Biggl({b \over b_0}\Biggr)^D D(D-1)g^{\mu\nu} {\p_{\mu}b
\p_{\nu}b \over b^2} \n \\
+ \Phi \b^D {D(D-1) \over b^2} + \o \b^D
g^{\mu\nu} {\p_{\mu} \Phi \p_{\nu} \Phi \over \Phi} + \b^D
\biggl({\cal L}(\s) - \L\biggr)\Biggr]
\end{eqnarray}
where
\be
\Phi = {\Omega_D b_0^D \over 16\pi} {\t \Phi} \n \\
\o = {{\t \o} \over \Omega_D} \n \\
\L = {\Omega_D b_0^D \over 16\pi} {\t \L}
\en
with $\Omega_D$ given by (3).

In this model we include the Casimir energy contribution which may
arise due to the compact nature of the internal space, which for a
$D$-sphere takes the form $A/b^{4+D}$ [26] where $A$ is a constant.
This term can provide the repulsive pressure at small values of the
internal scale factor $b(t)$ needed to balance the rolling down to
zero value of $b(t)$ [19,26]. The conformal Einstein frame can be
defined with the transformation
\be
{\bar g}_{\mu\nu} = {16\pi \Phi \over \m} \b^D g_{\mu\nu}
\en

The equations for motion in the Einstein frame have been solved
numerically~[19] where inflationary behaviour for the four
dimensional scale factor $a(t)$ is obtained together with stable
compactification of the internal manifold. The average time
variation of the higher dimensional JBD field $\Phi$ can be made
negligible with a suitable choice of parameters. It is required
that the total cosmological constant should vanish at the end of
the inflationary phase transition for emergence of the radiation
dominated era. The above desiradata enable the elimination of two
of the parameters in terms of the others, i.e.,
\be
A = - {8\pi \over \m} D(D-1) \Phi_0 b_0^{D+2} \n \\
\L = {D(D-1) \Phi_0 \over 2b_0^2}
\en
where the present value of $\Phi$ is taken as $\Phi_0$ which is
required to match the strength of the gravitational coupling, and
is thus given by
\be
\Phi_0 = {\m \over 16\pi}
\en
for this model. From (22) it can be seen that the bound $\o > 500$
ensures that the model stays within the present experimental limits
of allowed departure from general relativity after the internal
scale factor $b(t)$ settles down at the minimum of its potential.

In order to apply the other conditions for a viable EI scenario,
one has to consider the potential function for this model. In terms
of the dimensionless scalar field $y$ defined as
\be
y = D ln \b
\en
the Einstein frame potential (for $D=6$) is given by
\be
V(y) = \Biggl({\m V_0 \over 8\pi} + {15\Phi_0 \over b_0^2}\Biggr)
exp\biggl[-(y + 2\th)\biggr] - {15 \m \over 8\pi b_0^2}
exp\biggl[-({4\over 3}y + \th)\biggr] \n \\
+ {15 \Phi_0 \over b_0^2} exp\biggl[-({8\over 3}y + 2\th)\biggr]
\en
where $V_0$ is given in (3) and $\th = ln(16\pi\Phi / \m)$. The
potential for this model in terms of the dimensionless function
$W(y) = 16\pi V(y)/\m V_0$ is plotted in FIG.2. It was confirmed by
numerical integration of the equations of motion~[19]
that $y$ undergoes oscillations about the minimum of the potential
with decreasing amplitude. The contribution to the total energy
density by these oscillations is approximately constant in time,
thus aiding the inflationary behaviour of the scale factor $a(t)$.

To check the density perturbation constraint (19), we first
define the parameters $\e(y)$ and $\eta(y)$ (11) for this model.
It can be checked that the requirement of inflation ($\e , \eta
\sim 10^{-1}$) leads to the constraint
\be
V_0b_0^2 \simeq 10^{-2}
\en
for the present model. (Note that the relation (13) between the
number of e-foldings and $y$, which was derived for the potential
(8) of the model considered in Section II, is no longer valid in
the present case. Here the number of e-foldings is proportional to
the number of oscillations undertaken by $y$ before it settles down
at the minimum of its potential.) The density perturbation
constraint (19) can be satisfied if the quantity $y_{55}$ lies
within a narrow range of values. With the choice
(25), (26) and
(29) of the parameters, we find using (28) and (11) that to
generate appropriate density perturbations, $y$ should start from a
suitable initial value such that
\be
0.2 \le y_{55} \le 0.6
\en
Furthermore, (19) imposes an additional condition on the
behaviour of the $\Phi$ field, i.e.,
\be
{\Phi_{55} \over \Phi_0} \simeq 10^2
\en

Finally, let us examine the bubble spectrum of this model. Similar
to the case of the model of Section II, Here
again, the nucleation rate is time dependent~[17,18,19,24], since
in the dimensionally reduced action (22) the inflaton sector is
coupled to the internal scale factor $b(t)$ in the Jordan frame
itself. As in Section II, we consider the condition (16) which,
after some algebra, and upon substitution of the relevant
parameters translates into the constraint (for Einstein frame
quantities)
\be
{V_{55} \over V_{end}} > 10^5 exp\Biggl[-B_0\biggl(exp(y_{55}) -
1\biggr)\Biggr]
\en
Taking into account the fact that the range of allowed values of
$y_{55}$ from density perturbations (30) is already rather
narrow, it can be seen that (32) does not lead to any new
constraint on the parameters of the higher dimensional model. For a
typical choice of parameters, e.g., those used in plotting FIG.2,
one obtains $V_{55} \approx V_{end}$ and $y_{55} \approx 0.3$.
Substituting in (32), it is easy to see that the inequality holds
if the flat space bounce action $B_0 \ge 30$.

\section{Conclusions}

We have studied two kinds of higher dimensional models which lead
to JBD type theories upon dimensional reduction to four dimensions.
The numerical solutions of these models had been worked out earlier
[17,19] which showed that enough inflation together with dynamical
compactification of the extra dimensions is possible, thus
bolstering their {\it a priori} feasability as
candidates for EI. A
crucial common feature of these models is that the time dependent
scale factor of the internal manifold is coupled to the inflaton
sector of the four dimensional effective action in the Jordan
frame. This causes the nucleation rate of the true vacuum bubbles
to be time dependent~[17,18,19,24]. Furthermore, this feature leads
to a different form for the effective inflaton potential and the
corresponding slow role parameters in the conformal Einstein frame,
from that obtained in the usual models of EI~[6,11,12,13]. The
models considered by us, thus clearly lie outside the ambit of the
general class of models examined in [16].

In this paper we have applied to the framework of these models four
essential conditions~[16] required for the viability of any EI
model, which are (a) recovering general relativity within the
present day experimental limits, (b) reproducing the present value
of the gravitational coupling, (c) producing a bubble spectrum
conforming to CMBR isotropy~[25], and (d) generating density
perturbations compatible with present large scale structure
observations~[15]. The model considered in Section II is able to
satisfy the first three of these desiradata which in turn impose
the constraints (9), (10) and (17) on the parameters used.
However, it fails to meet the criterion of appropriate density
perturbations. The latter model (Section III) can be made
compatible with all the above conditions, albeit only for  a narrow
range of parameters (25), (26), and (29)---(32). Our analysis
shows that the simpler higher dimensional models~[17,18] are not
viable as candidates of successful EI, although more complicated
ones~[19] (e.g., with the inclusion of extra fields, and quantum
effects in the lagrangian) are just able to squeze through the
conditions imposed by present observations through stringent
constraints on parameters.

\vskip 1.0in

The author wishes to acknowledge the financial support provided
through a project by
the Department of Science and Technology, Government of India.

\pagebreak

REFERENCES

\vskip 0.2in

\begin{description}

\item[[1]] See, for example, ``Modern Kaluza-Klein
theories'', edited by T.Appelquist, A.Chodos and
P.G.O.Freund (Addison-Wesley, Reading, MA, 1987).
\item[[2]] P.G.O.Freund, Nucl. Phys. B {\bf 209},
146 (1982); P.G.O.Freund and P.Oh, {\it ibid}.
{\bf 255}, 688 (1985); G.F.Chapline and
G.W.Gibbons, Phys. Lett. B {\bf 135}, 43 (1984);
K.Maeda, {\it ibid}. {\bf 138}, 269 (1984);
{\it ibid}. {\bf 166}, 59 (1986); D.Bailin and
A.Love, {\it ibid}. {\bf 163}, 135 (1985); {\it
ibid}. {\bf 165}, 270 (1985); A.S.Majumdar,
A.Mukherjee and R.P.Saxena, Mod. Phys. Lett. A
{\bf 7}, 3647 (1992).
\item[[3]] I.Antoniadis, C.Bachas, J.Ellis and
D.V.Nanopoulos, Phys. Lett. B {\bf 211}, 393
(1988); Nucl. Phys. B {\bf 328}, 117 (1989); Phys.
Lett. B {\bf 257}, 278 (1981); M.Mueller, Nucl.
Phys. B {\bf 337}, 37 (1990); B.A.Campbell,
M.J.Duncan, N.Kaloper and K.A.Olive, {\it ibid}.
{\bf 351}, 779 (1991); A.A.Tseytlin, Int. Jour.
Mod. Phys. D {\bf 1}, 223 (1992); A.A.Tseytlin and
C.Vafa, Nucl. Phys. B {\bf 372}, 443 (1992);
R.Brustein and P.J.Steinhardt, Phys. Lett. B {\bf
302}, 196 (1993).
\item[[4]] P.G.Bergman, Int. J. Teor. Phys. {\bf
1}, 25 (1968); R.V.Wagoner, Phys. Rev. D {\bf 1},
3209 (1970); K.Nordtvedt, Ap. J. {\bf 161}, 1059
(1970).
\item[[5]] P.Jordan, Z. Phys. {\bf 157}, 112
(1959); C.Brans and R.H.Dicke, Phys. Rev. {\bf
124}, 925 (1961).
\item[[6]] C.Mathiazhagan and V.B.Johri, Class
Quant. Grav. {\bf 1}, L29 (1984); D.La and
P.J.Steinhardt, Phys. Rev. Lett. {\bf 62}, 376
(1989); E.W.Kolb, Phys. Scr. {\bf 36}, 199 (1991).
\item[[7]] A.H.Guth, Phys. Rev. D {\bf 23}, 347
(1981).
\item[[8]] See, for instance, A.D.Linde,
``Particle Physics and Inflationary Cosmology'',
(Harwood Academic, Chur, Switzerland, 1990);
E.W.Kolb and M.S.Turner, ``The Early Universe'',
(Addison-Wesley, Redwood City, CA, 1990);
A.R.Liddle and D.H.Lyth, Phys. Rep. {\bf 231}, 1
(1993).
\item[[9]] A.Albert and P.Steinhardt, Phys. Rev.
Lett. {\bf 48}, 1220 (1982); A.D.Linde, Phys.
Lett. B {\bf 129}, 177 (1983).
\item[[10]] R.D.Reasenberg, et al., Ap. J. {\bf
234}, L219 (1979); C.M.Will, ``Theory and
Experiment in Gravitational Physics'', (Cambridge
University Press, Cambridge, 1993).
\item[[11]] E.J.Weinberg, Phys. Rev. D {\bf 40},
3950 (1989); D.La, P.J.Steinhardt, and
E.Bertschinger, Phys. Lett. B {\bf 231}, 231
(1989).
\item[[12]] R.Holman, E.Kolb and Y.Wang, Phys.
Rev. Lett. {\bf 65}, 17 (1990); P.J.Steinhardt and
F.S.Accetta, Phys. Rev. Lett. {\bf 64}, 274
(1990); R.Crittenden and P.J.Steinhardt, Phys.
Lett. B {\bf 293}, 32 (1992); A.Layock and
A.R.Liddle, Phys. Rev. D {\bf 49}, 1827 (1994).
\item[[13]] F.S.Accetta and J.J.Trester, Phys.
Rev. D {\bf 39}, 2854 (1989); A.D.Linde, Phys.
Lett. B {\bf 249}, 18 (1990); A.Burd and A.Coley,
{\it ibid}. {\bf 267}, 330 (1991); J.D.Barrow and
K.Maeda, Nucl. Phys. B {\bf 341}, 294 (1991);
A.R.Liddle and D.Wands, Phys. Rev. D {\bf 45},
2665 (1992); F.Occhionero and L.Amendola, Phys.
Rev. D {\bf 50}, 4846 (1994).
\item[[14]] G.F.Smoot et al., Ap. J. {\bf 396}, L1
(1992).
\item[[15]] A.R.Liddle and D.H.Lyth, Phys. Lett. B
{\bf 291}, 391 (1992); Ann. N.Y. Acad. sci. {\bf
688}, 653 (1993); J.Garcia-Bellido and D.Wands,
Phys. Rev. D {\bf 52}, 6739 (1995); A.R.Liddle,
D.H.Lyth, R.K.Shaeffer, Q.Shafi and P.T.P.Viana,
Mon. Not. R. Astron. Soc. {\bf 281}, 531 (1996).
\item[[16]] A.M.Green and A.R.Liddle, Phys. Rev. D
{\bf 54}, 2557 (1996).
\item[[17]] A.S.Majumdar and S.K.Sethi, Phys. Rev.
D {\bf 46}, 5315 (1992).
\item[[18]] R.Holman, E.W.Kolb, S.L.Vadas and
Y.Wang, Phys. Rev. D {\bf 43}, 995 (1991).
\item[[19]] A.S.Majumdar, T.R.Seshadri and
S.K.Sethi, Phys. Lett. B {\bf 312}, 67 (1993).
\item[[20]] Y.M.Cho, Phys. Rev. Lett. {\bf 68},
3133 (1992).
\item[[21]] K.Sunahara, M.Kasai and T.Futamase,
Prog. Theor. Phys. {\bf 83}, 353 (1990).
\item[[22]] K.Maeda, Phys. Rev. D {\bf 39}, 3159
(1989).
\item[[23]] F.S.Accetta and P.Romanelli, Phys.
Rev. D {\bf 41}, 3024 (1990).
\item[[24]] R.Holman, E.W.Kolb, S.L.Vadas, Y.Wang
and E.J.Weinberg, Phys. Lett. B {\bf 237}, 37
(1990); R.Holman, E.W.Kolb, S.L.Vadas, and Y.Wang,
{\it ibid}. {\bf 250}, 24 (1990).
\item[[25]] A.H.Guth and E.J.Weinberg, Nucl. Phys.
B {\bf 212}, 321 (1983); E.J.Copeland, A.R.Liddle,
D.H.Lyth, E.D.Stewart and D.Wands, Phys. Rev. D
{\bf 49}, 6410 (1994).
\item[[26]] F.S.Accetta, M.Gleiser, R.Holman and
E.W.Kolb, Nucl. Phys. B {\bf 276}, 501 (1986);
A.D.Linde and M.I.Zelnikov, Phys. Lett. B {\bf
215}, 59 (1988); L.Amendola, E.W.Kolb, M.Litterio
and F.Occhionero, Phys. Rev. D {\bf 42}, 1944
(1990).

\end{description}

\pagebreak

FIGURE CAPTIONS

\vskip 0.5in

{\bf Fig.1} \hskip 0.5in The potential for the
model with nonminimal inflaton coupling is plotted
versus the internal scale factor which has the
form of a JBD field in four dimensions.

\vskip 0.2in

{\bf Fig.2} \hskip 0.5in The potential for the
model of Sec.III shows that the JBD field $y$ can
undergo oscillations before settling down at the
minimum. For the values of parameters used in this
plot, the choice of $y_{55} \approx 0.3$ satisfies
the density perturbation constraints.

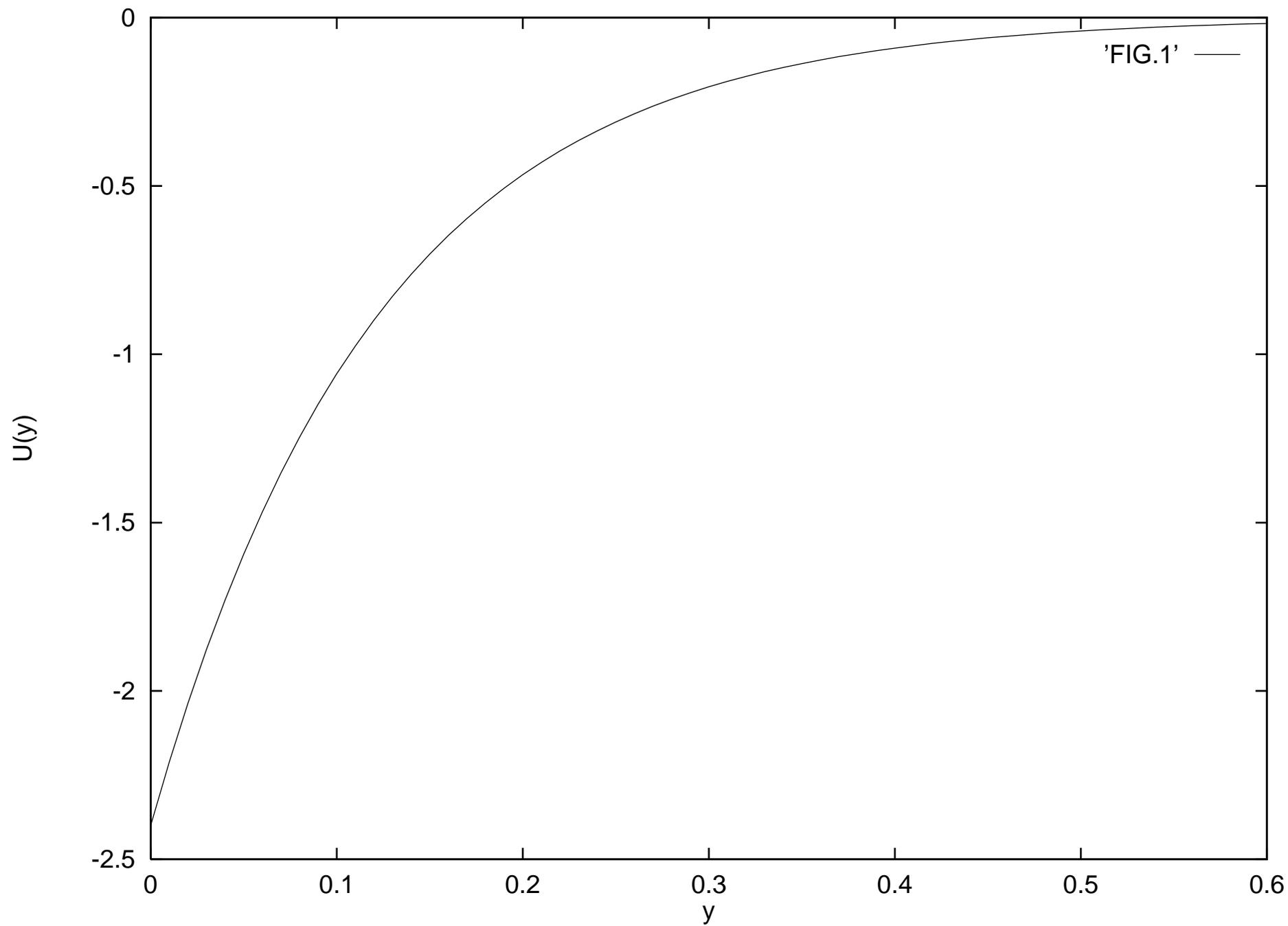
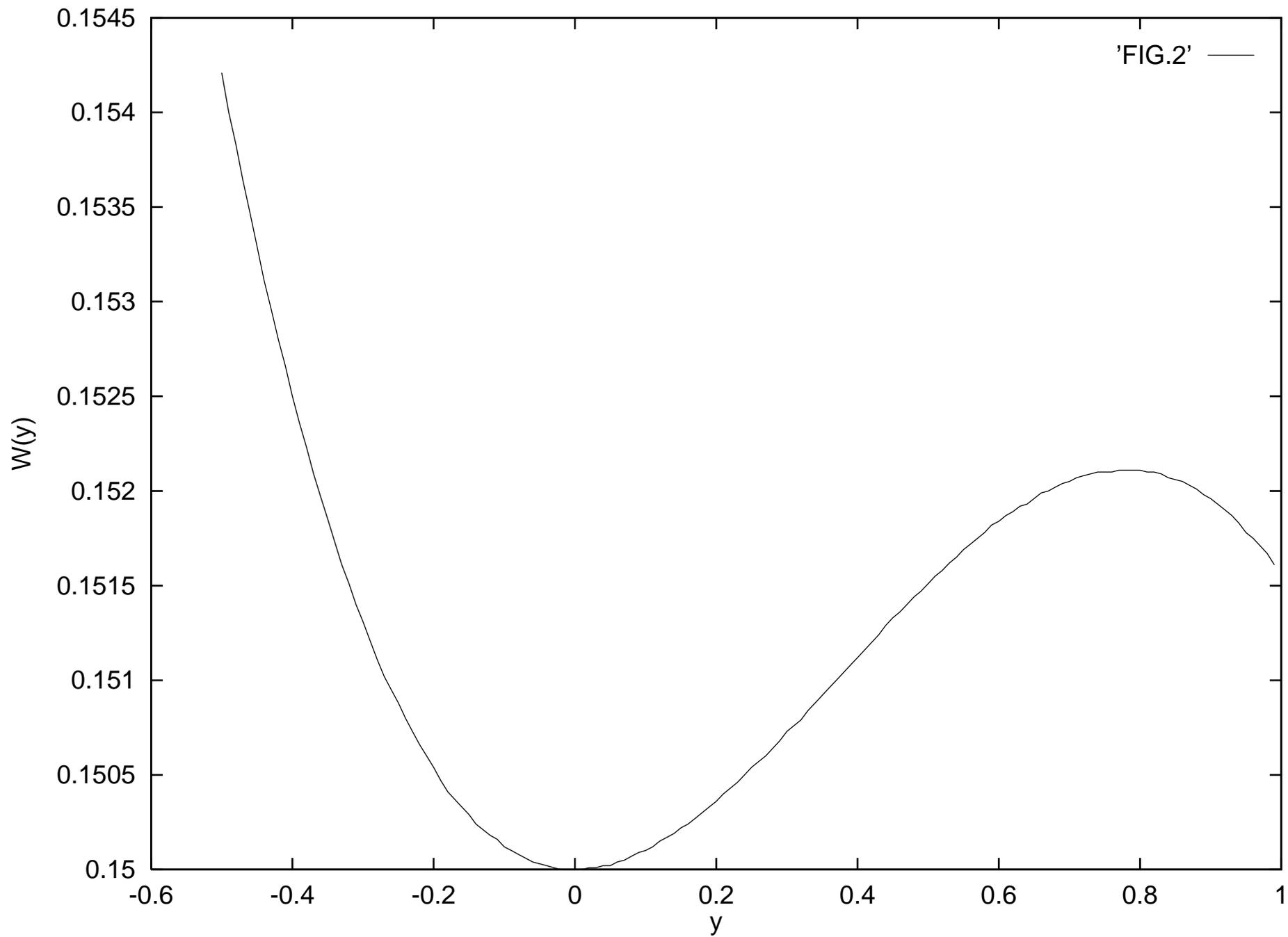
\end{document}